\documentclass[sigconf]{acmart}
\def\BibTeX{{\rm B\kern-.05em{\sc i\kern-.025em b}\kern-.08emT\kern-.1667em\lower.7ex\hbox{E}\kern-.125emX}}
\usepackage{bm}
\usepackage{subfigure}
\copyrightyear{2019}
\acmYear{2019}
\setcopyright{acmlicensed}
\acmConference[MAISoN '19]{MAISoN '19: 3rd International Workshop on
Mining Actionable Insights from Social Networks}{October 02--05, 2019}{Santa Clara, CA}
\acmBooktitle{MAISoN '19: MAISoN '19: 3rd International Workshop on
Mining Actionable Insights from Social Networks, October 02--05, 2019, Santa Clara, CA}

\begin{document}

%
% The "title" command has an optional parameter, allowing the author to define a "short title" to be used in page headers.
\title{Convolutional Quantum-Like Language Model with Mutual-Attention for Product Rating Prediction}

%
% The "author" command and its associated commands are used to define the authors and their affiliations.
% Of note is the shared affiliation of the first two authors, and the "authornote" and "authornotemark" commands
% used to denote shared contribution to the research.

\author{Qing Ping}
\email{qp27@drexel.edu}

\affiliation{%
  \institution{Drexel Univeristy}
  \streetaddress{3141 Chestnut St}
  \city{Philadelphia}
  \state{PA}
  \postcode{95104}
}

\author{Chaomei Chen}
\email{cc345@drexel.edu}
\affiliation{%
\institution{Drexel Univeristy}
  \streetaddress{3141 Chestnut St}
  \city{Philadelphia}
  \state{PA}
  \postcode{95104}
  }

%
% By default, the full list of authors will be used in the page headers. Often, this list is too long, and will overlap
% other information printed in the page headers. This command allows the author to define a more concise list
% of authors' names for this purpose.

%
% The abstract is a short summary of the work to be presented in the article.

\begin{abstract}
Recommender systems are designed to help mitigate information overload users experience during online shopping. Recent work explores neural language models to learn user and item representations from user reviews and combines such representations with rating information. Most existing convolutional-based neural models take pooling immediately after convolution, and loses the interaction information between latent dimension of convolutional feature vectors along the way. Moreover, these models usually take all feature vectors at higher levels as equal, and do not take into consideration that some features are more relevant to this specific user-item context. To bridge these gaps, this paper proposes a convolutional quantum-like language model with mutual-attention for rating prediction (\textbf{ConQAR}). By introducing a quantum-like density matrix layer, interactions between latent dimensions of convolutional feature vectors are well captured. With the attention weights learned from the mutual-attention layer, final representations of a user and an item absorb information from both itself and its counterparts for making rating prediction. Experiments on two large datasets show that our model outperforms multiple state-of-the-art CNN-based models. We also perform ablation test to analyze the independent effects of the two components of our model. Moreover, we conduct a case study and present visualizations of the quantum probabilistic distributions in one user and one item review document to show that the learned distributions capture meaningful information about this user and item, and can be potentially used as textual profiling of the user and item.
\end{abstract}




%
% Keywords. The author(s) should pick words that accurately describe the work being
% presented. Separate the keywords with commas.

%
% A "teaser" image appears between the author and affiliation information and the body 
% of the document, and typically spans the page. 

%
% This command processes the author and affiliation and title information and builds
% the first part of the formatted document.
\maketitle
\section{Introduction}
With the booming of the e-commerce industry, users enjoy ever-efficient shopping experience with any desired product one-click purchase away. In the meanwhile, users are faced with millions of indistinguishable products and suffer from cognitive information overload when making decisions. Recommender systems are designed to help mitigate such information overload by recommending a subset of products that most likely users may like.

Collaborative filtering-based approaches have shown great performance over large datasets where only rating information is available. Yet when such dataset is small, the performance usually deteriorates, known as the "cold-start" problem. In the meanwhile, more and more user review information becomes available nowadays, presenting new opportunities and challenges in the research topic of incorporating textual information into rating prediction.

On the one hand, user reviews may contain extra information about a user's preferences. If distilled properly, such information may be beneficial for improving product rating prediction. On the other hand, incorporating textual information into rating prediction is non-trivial, where in extreme cases, the fusion of the two can even hurt model performance.

Earlier work along this line combined topic modeling with collaborative filtering \cite{mcauley2013hidden,diao2014jointly,ling2014ratings}. These methods have shown potentials to improve rating prediction performance compared with rating-based collaborative filtering methods. However, such models also suffer from similar issues as topic models, such as loss of sentence structure and sparsity of semantics in short review documents. Later research turns to neural language models \cite{zheng2017joint,seo2017representation,catherine2017transnets,chen2018neural,ni2017estimating,li2017neural}, where convolutional neural nets (CNNs) \cite{zheng2017joint,seo2017representation,catherine2017transnets,chen2018neural} and  recurrent neural nets (RNNs) \cite{ni2017estimating,li2017neural} are explored to learn textual representation of user reviews, and combine with neural collaborative filtering networks. Here, how text is represented, and how to merge such information from a user and an item into one final representation, are two important research questions. For CNN-based approaches, most existing models utilize one-dimensional convolutional layers followed by max or mean pooling as representations of users and items, during which, we argue that the interactions between latent dimensions of feature vectors are lost due to immediate pooling. Moreover, most existing approaches merge user and item representations with vector concatenation, vector dot product or factorization machine, where all features are taken equal, and features that exhibit strong signals in both user and item review do not stand out.

To bridge these gaps, the present study propose a \textbf{conv}olutional \textbf{q}uantum-like language model with mutual-\textbf{a}ttention for \textbf{r}ating prediction (\textbf{ConvQAR}). The main components of the model are convolutional layers, quantum density matrix layers, and mutual-attention layers. The quantum density matrix layers borrow an analogy from quantum event representations, and learns the interactions between latent dimensions of a convolutional feature vector. The mutual-attention layer is designed in such a way that features that exhibit stronger signals in both user and item will be given higher weight in final representation, therefore making the representations context-aware with the designed communicative mechanism. 

The contributions of the present paper are as follows:
(1) we propose a quantum-like density matrix layer to learn interactions in latent dimensions of feature vectors that are neglected in previous CNN-based models;
(2) we propose a mutual-attention layer to communicate information between a user and an item, so that final representation is weighted towards feature vectors that exhibit stronger signals on both sides;
(3) we explore the internal mechanism of the density matrix layer by visualizing dyad probabilistic distributions over user and item reviews. 

\section{Related Work}
\subsection{Topic Modeling-based Approach}
Early work represents review documents with topic models, and aligns the topic distributions with rating matrices. HFT aligns hidden topics in product reviews with latent dimensions in product ratings, where topics serve as regularisers of latent user
and product matrices \cite{mcauley2013hidden}. JMARS aligns the aspects identified in reviews with hidden preferences of users and item from rating matrices, and predict sentiments on each aspect \cite{diao2014jointly}. RMR avoids the difficult choice of the transformation function
in HFT and retain the interpretability of the latent topics with a mixture of Gaussian for ratings \cite{ling2014ratings}.

All these models have improved model performance compared with pure rating-based approaches. However, they also suffer from similar issues as topic models. First, when dataset is small and reviews are short, these models have to estimate topic distributions from very sparse semantics. Second, these models do not consider sentence structures or n-grams, which may lead to inaccurate representations of the review documents.  

\subsection{Neural Language Model Approach}
Some later work utilizes convolutional neural networks (CNNs) to learn abstract representations of the user and item review documents, and combine such representations with rating information. More specifically, DeepCoCNN uses two parallel CNNs to represent each user and item, and use factorization machine at later stage to learn interactions between the two \cite{zheng2017joint}. Dual-Att further improves DeepCoCNN by introducing a global and local attention layer on top of the CNN layers \cite{seo2017representation}. TransNets designed a teacher-student network to force the merged representations of a user and item to approximate the representation of the corresponding review for this pair of user and item \cite{catherine2017transnets}. NARRE further incorporates usefulness information into the model, and introduces an attention layer to learn attentions of features based on user and item embeddings \cite{chen2018neural}.

As emphasized in Introduction, most existing CNN-based models utilize one-dimensional convolutional layers followed by max or mean pooling as representations of users and items. The interactions between latent dimensions of feature vectors are lost along the way. Moreover, most existing approaches merge user and item representations with vector concatenation, vector dot product or factorization machine, where all features are taken equal, and features that exhibit strong signals in both user and item review do not stand out.

There is also some work that utilizes RNN models for rating prediction \cite{ni2017estimating,li2017neural}. Since the present paper focuses on improving CNN-based language models, we will not further elaborate on this line of works. 
\subsection{Quantum-Like Language Models}
Pioneer work \cite{sordoni2013modeling} propose to model term dependencies with quantum-like two-dimensional representations instead of traditional one-dimensional representation. Moreover, the probabilities for the density matrices are estimated with R$\rho$R EM algorithm.

In later work for question answering \cite{zhang2018end}, an end-to-end quantum-like language model is proposed, where probabilities are not estimated with Expectation-Maximization algorithm, but learnt by the neural model in end-to-end fashion. Also, dense word embeddings are used, instead of one-hot encodings. 

The differences between our model and previous work \cite{zhang2018end} are two-folds. First, we perform quantum-like dyads on top of convolutional features instead of word embeddings, therefore the basic events here are n-grams instead of words. Second, we use the max row-pooling and column-pooling of interaction matrix as attentions for weighting user and item density matrices, instead of using only the interaction matrix itself for deriving final representations.

The present paper borrows analogous ideas from quantum theory and previous quantum language models. Concepts and ideas are discussed in section 3.

\section{Preliminaries of Quantum Language Model}
\subsection{Basic Concepts}
\textbf{Hilbert space} $\mathbb{H}^n$: In quantum theory, the probabilistic space is naturally represented as a vector space, or Hilbert space $\mathbb{H}^n$. This space is simplified to be finite real space $\mathbb{R}^n$, same as in previous work \cite{sordoni2013modeling}. $\textbf{ket}\;|\textbf{u}\rangle$ and $ \textbf{bra}\;\langle \textbf{u}|$: a unit vector $\bm{u}\in\mathbb{R}^n,{||\bm{u}||}_2=1$ is written as a $ket\;|\bm{u}\rangle$ under Dirac's notation of real spaces, and its transpose $\bm{u}^T$ is written as a $bra\;\langle \bm{u}|$.  \textbf{Projectors}: a projector of a state $\bm{u}$ onto a vector $\bm{u}$ is denoted as  $|\bm{u}\rangle\langle \bm{u}|$.  \textbf{Events}: unlike in classic probability theory, where events are subsets, events in quantum theory are defined as subspaces,  or projectors onto subspaces   \cite{nielsen2002quantum,park2011quasi}. \textbf{Basic events}(dyad): a state vector $\bm{u}$ projected onto the unit norm vector itself $\bm{u}\in\mathbb{R}^n,{||\bm{u}||}_2=1$ is an elementary event, denoted as $|\bm{u}\rangle\langle \bm{u}|$. \textbf{Density matrix}: a density matrix can be seen as representation of an event, namely a mixture of elementary events (or dyads): 
\begin{equation}
\rho= {\sum_{i} p_i|\psi_i\rangle\langle \psi_i|}
\label{eq:1}
\end{equation}
where $|\psi_i\rangle$ is a unit state vector with probability $p_i$. The density matrix $\rho$ has several properties, (1) symmetric, (2)positive semi-definite, and (3) has trace of 1. By Glenson's theorem, a density matrix assigns a quantum probability to each one of the infinite dyads, and can be seen as a quantum generalization of a classical probability distribution.

\subsection{Quantum Language Model}
A quantum language model represents each word or compound dependency between words with a quantum elementary event. In earlier work\cite{sordoni2013modeling}, each word $w_i$ is associated with a projector  dyad $|e_i\rangle\langle e_i|$, where $e_i$ a standard basis vector, or one-hot vector. To calculate the similarity between a query and a document, the density matrix $\rho_q$ of a query and the density matrix $\rho_d$ of a document were estimated with a R$\rho$R algorithm, which iteratively updates the density matrices through Expectation Maximization (EM) algorithm. After the estimation of density matrices, the final ranking is based on Von-Neumann (VN) Divergence between $\rho_q$ and $\rho_d$. Since there is no analytical solution to the EM algorithm, this quantum language model cannot be integrated to an end-to-end deep language model framework.

In later work on question answering \cite{zhang2018end}, instead of using one-hot encoding for each word, normalized word embedding trained on global corpus was used. Also, instead of estimating density matrices with the R$\rho$R EM algorithm, the probabilities $p_i$ is left for the model to learn in an end-to-end neural network fashion. Once $p_i$ is estimated, the density matrix of a question or an answer can be derived with equation \ref{eq:1}. 

\begin{figure}
\includegraphics[width = 0.5\textwidth]{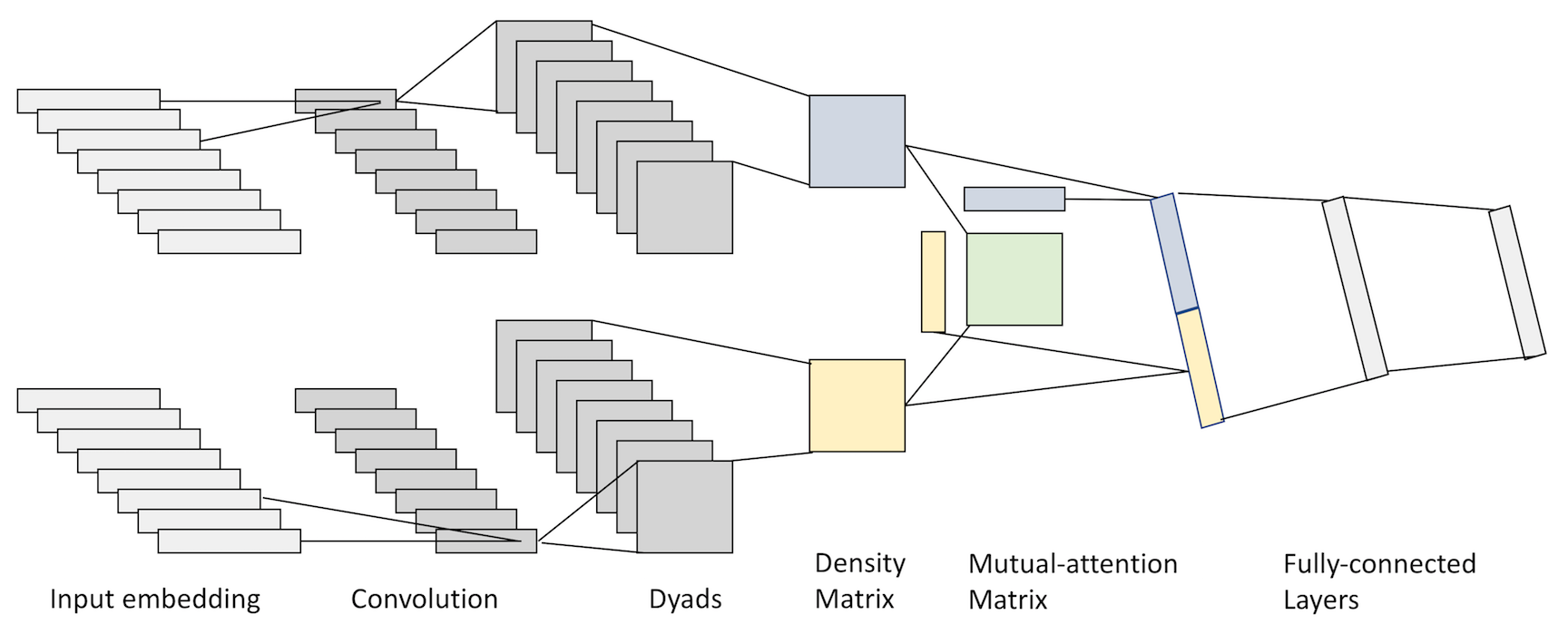}
\caption{Structure of Proposed ConQAR Model.} \label{fig1}
\end{figure}

\section{Convolutional Quantum Language Model }
 The high-level idea of our model is that (1) instead of using one-dimensional convolutional network as representation of reviews of each user and item, we add a quantum-like density matrix representation on top of the convolutional layer to bring in extra interaction information between latent features; (2) instead of learning user and item representation in parallel and using dot product or factorization machine to learn their interaction at final stage, we introduce a mutual-attention layer to attend to each density matrix that absorbs information from counterparts. 
 In following sections, we describe the details in our model.
\subsection{Embedding Layer}
The overall structure of our proposed model is depicted in Fig.~\ref{fig1}. We represent each user $u$ and each item $v$ with an embedding layer that performs lookup functions from a set of one-hot encoding vectors $(\vec{e_1},\vec{e_2},...,\vec{e_L} ),\vec{e_i} \in \mathbb{R}^{|V|}$. Each one-hot encoding represents a word in review document. The lookup function maps words to a set of dense vectors $(\vec{x_1},\vec{x_2},…,\vec{x_L} ),\vec{x_i}\in\mathbb{R}^d$, based on weight matrix $\vec{W}_e\in\mathbb{R}^{d\times|V|}: \vec{x_i}=\vec{W}_e\cdot \vec{e_i} $. The dense vectors are then concatenated as an embedding matrix for each review document: $\vec{X_{1:L}}=\vec{x_1}\oplus \vec{x_2}\oplus ... \oplus \vec{x_L}$, where $V$ is the vocabulary of the corpus, $d$ is the dimension of word embedding, and $L$ is the length of a review document. 

\subsection{Convolutional Layer}
A convolution operation applies a filter $\vec{w}_{conv}\in \mathbb{R}^{d\times h}$ to the word embedding matrix $\vec{X}$ with a window size $h$, derives a feature value $c_i$ from such convolution operation.
\begin{equation}
c_i=f(\vec{w}_{conv} \vec{x}_{i:i+h-1}+b)
\end{equation}
Where $f$ is a non-linear function such as RELU\cite{nair2010rectified}, $b$ is the bias term, and $h$ is the size of the filter.
By sliding the filter window one-dimensionally across different word positions in a review document, we can derive a feature vector $\vec{c}_s$ for each filter $s$:
\begin{equation}
\textbf{c}(s)=[c_1,c_2,...,c_{L-h+1}]
\end{equation}
Then all $n$ feature vectors all of different window sizes comprise the final feature map $\vec{C} \in \mathbb{R}^{n \times L}$ :
\begin{equation}
\textbf{C}={[\vec{c}(s_1),\vec{c}(s_2),...,\vec{c}(s_n)]}^T
\end{equation}

\subsection{Quantum Density Matrix Layer}
Formally, convolutional feature map $\textbf{C} \in \mathbb{R}^{n \times L}$ encodes the n-gram semantic features of a review document. Such a distributed representation of each n-gram feature can naturally serve as an observed state for a review document. The entire document can be regarded as a mixture of these n-gram states. Compared to one-hot encoding of single words, the convolutional feature maps of n-grams are distributional dense vectors and capture word co-occurrence relationships since it is derived from word embeddings trained a large global corpus. Compared with single-word distributional embeddings, it is semantically richer since it captures frequent n-gram contexts around words instead of words alone. 

To obtain a unit state vector, we normalize each convolutional feature vector as:
\begin{equation}
|c\rangle= \frac {\vec{c}}{||\vec{c}||_2}
\end{equation}
Each review document correspond to a mixture of unit state vectors namely a density matrix, which is defined as:
\begin{equation}
\rho=  {\sum\limits_{i} p_i|c\rangle\langle c|}
\end{equation}
Where $\rho$ is the density matrix that is symmetric, positive semidefinite
and has trace of 1. $|c\rangle\langle c|$ is the semantic subspace spanned on the feature unit state vector $\vec{c}$ of convolutional operations. $\vec{p}=(p_1,p_2,p_3,...), \sum p_i=1$ is the corresponding probability distribution, where each $p_i$ corresponds to  a unit state $|c_i\rangle$ in a review document.

In practice, the values of $\vec{p}$ reflect the weights of different n-grams of a document. It can be considered as parameters that can be learned in our end-to-end model. Moreover, since the entire space of all n-gram possibilities is computationally large, we approximate the n-gram distribution with a probabilistic distribution over different positions of a document, and assume a distinct distribution for each user and item. In this sense, our model is not a strictly-speaking quantum model but an analogous quantum-like model that takes advantage of useful properties of density matrices. 

The dyads $|c\rangle\langle c|$ in density matrices can also be interpreted as a reflection of covariance of latent dimensions in an n-gram feature vector in the convolutional feature maps. In other words, it represents how scattered the n-gram features will be in the latent feature space. This information is not captured in one-dimensional convolutions.

To satisfy the trace=1 condition for the density matrices, we add a loss to enforce that the sum of diagonal elements in density matrices is close to 1:
\begin{equation}
L_{Trace}= \frac{1}{U}\sum\limits_{u \in users} (\sum_{i=1}^{n}r_{ii}(u) - 1)^2  + \frac{1}{V}\sum\limits_{ v \in items} (\sum_{i=1}^{n}r_{ii}(v) - 1)^2 
\end{equation}
where $r_{ii}(u)$ and $r_{ii}(v)$ represent the $ith$ diagonal element in the density matrix ${\rho}_u$ of user $u$ and ${\rho}_v$ of item $v$, and $U$ and $V$ represent the total number of users and items.

\subsection{Mutual Attention Layer}
Given user $u$ represented as two-dimensional density matrix $\rho_u$ and item $v$ as $\rho_v$ respectively, we propose a way to learn the joint-representation of the two density matrices.

First, we take matrix multiplication of $\rho_u$ and $\rho_v$ as the mutual-attention matrix, inspired by earlier work \cite{zhang2018end}.

\begin{equation}
\textbf{M}_{uv}= \rho_u \cdot {\rho_v}^T
\end{equation}
Then  trace of $\textbf{M}_{uv}$ is an approximate measure of similarity between the two density matrices. 
\begin{equation}
tr(\rho_u \rho_v) = tr(\sum\limits_{i} p_i|c_i\rangle\langle c_i| \cdot \sum\limits_{j} p_j|c_j\rangle\langle c_j| )  = \sum\limits_{i,j} p_ip_j {\langle c_i| c_j\rangle}^2
\end{equation}
For details of the inference, please refer to previous work \cite{zhang2018end}. In short, the trace of $\textbf{M}_{uv}$ is a generalization of inner product of vectors to matrices. We will include this trace $tr(\rho_u \rho_v)$ and the diagonal elements $diag(\textbf{M}_{uv}) = [r_{11},r_{22},...,r_{nn}]$ as part of the final representation of a use and an item similar as in \cite{zhang2018end}.

Moreover, given the mutual-attention matrix $\textbf{M}_{uv}$ for each pair of user $u$ and item $v$, we perform row and column average pooling to derive the weighting vector $g_u$ for user $u$ and weighting vector $g_v$ for item $p$.
\begin{equation}
\vec{g}_u = [\theta_1,\theta_2,...,\theta_n], \theta_i= \frac{1}{n}\sum_{j=1}^{n} r_{ij}, r_{ij}\in \textbf{M}_{uv}
\end{equation}
\begin{equation}
\vec{g}_v = {[\gamma_1,\gamma_2,...,\gamma_n]}^T, \gamma_j= \frac{1}{n}\sum_{i=1}^{n} r_{ij}, r_{ij}\in \textbf{M}_{uv}
\end{equation}
Then we derive the mutual-attention vector $\vec{a}_i(u)$ of a user $u$ and $\vec{a}_i(v)$ of an item $v$ with a softmax layer:
\begin{equation}
\textbf{a}_i(u) = \frac{e^{\theta_i}}{\sum_{k=1}^{n} e^{\theta_k}}
\end{equation}
\begin{equation}
\textbf{a}_j(v) = \frac{e^{\gamma_j}}{\sum_{k=1}^{n} e^{\gamma_k}}
\end{equation}
The intuition behind the mutual-attention vectors is to capture the high-level information that are mutually shared in user and item text documents.  The $n \times n$ density matrices of a user and an item are an abstract representation of a document on the $n$ filters (n-gram features) space. By taking the mean row pooling from the mutual-attention matrix, we derive a weight vector to denote which n-gram feature of a user has greater values in the mutual-attention matrix, which is an approximate that this n-gram feature is averagely more similar to all n-grams in the document of the counterpart item, and vice versa.  

Then we compute the final representation of a user and an item by weighting the density matrices with the mutual-attention vector: \begin{equation}
\textbf{z}_u = {({\rho}_u \cdot \textbf{a}_u)}^T,  \textbf{z}_v = \textbf{a}_v \cdot {\rho}_v 
\end{equation}
Eventually, we concatenate the following parts into one final representation of a user and an item:
\begin{equation}
\textbf{z}_{uv} = tr(\rho_u \rho_v) \oplus  diag(\textbf{M}_{uv}) \oplus \textbf{z}_u  \oplus  \textbf{z}_v
\end{equation}

\subsection{Fully Connected Layer}
We feed the final representation $\textbf{z}_{uv}$ into $m$ fully-connected layers, where the first $m-1$ layers are defined as follows:
\begin{equation}
\vec{z}_{i} = f_i(\vec{W}_i \cdot \vec{z}_{i-1} + \vec{b}_i)
\end{equation}
where $\vec{W}_i$ and $\vec{b}_i$ are the weight matrix and bias vector of layer $i$, and $f_i(\cdot)$ is the non-linear ELU activation function \cite{clevert2015fast}. Dropout layers are applied after each layer.  

The final layer is similar to previous layers but without a non-linear activation function:
\begin{equation}
\hat{y}_{uv} = \vec{W}_m \cdot \vec{z}_{m-1} + b_m
\end{equation}
Then we define the mean-square error loss between this predicted rating $\hat{y}$  and ground-truth rating $y$ as our prediction loss:
\begin{equation}
L_{Rating}= \frac{1}{N}\sum_{i=1}^{N} {(\hat{y}_{i} - y_{i}  )}^2
\end{equation}
where $N$ is the total number of samples of user and item pairs.

\subsection{Final Loss}
The final loss of the model is a weighted sum of the losses mentioned before:
\begin{equation}
L_{ConQAR} = \alpha \cdot L_{Trace} + (1-\alpha)  \cdot L_{Rating}
\end{equation}

\section{Experiment}
In this section, we describe the details of our experiments, including datasets, evaluation metrics, baselines, and experimental results.
\subsection{Datasets}
We use two publicly available datasets for evaluation, namely Amazon Review Dataset and Yelp Dataset. Both datasets contain information of user, item, and corresponding rating and review of each pair of user and item. For the Amazon Review Dataset, we use the 5-core datasets since the ``cold-start'' problem is not the focus of this paper. Meanwhile, we select 7 product categories for comparison among all products in the Amazon Review Dataset. The Yelp Dataset is from the Yelp Business Rating Prediction Challenge in 2013 RecSys, which covers user reviews of restaurants in the Phoenix, AZ metropolitan area. Basic statistics of the datasets are listed in Table~\ref{tab1}.

\begin{table*}
  \caption{Basic Statistics of Amazon Review Dataset and Yelp Dataset.}
  \label{tab1}
  \begin{tabular}{p{1.3cm}p{1.3cm}p{1.3cm}p{1.3cm}p{1.3cm}p{1.3cm}p{1.3cm}p{1.3cm}p{1.3cm}}
    \toprule
    &Musical Instruments&Automo-tive&Amazon Instant Video&Office Products&Patio Lawn and Garden&Grocery and Gourmet Food&Digital Music & Yelp\\
    \midrule
    reviews&10,261&20,473&37,126&53,258&13,272&151,254&64,706&252,897\\
users &1,429&2,928&5,130&4,905&1,686&14,681&5,541&45,981\\
items &900&1,835&1,685&2,420&962&8,713&3,568&11,537\\
    \bottomrule
  \end{tabular}
\end{table*}

\subsection{Data Preprocessing}

To represent each user and item, we concatenate all reviews of a user (item) into one text document as a representation of the user (item). In test dataset, we exclude the particular review among all reviews of representation for a particular user (item) to avoid ground-truth leakage. A delimiter is added between every two consecutive sentences and each concatenated two reviews. The maximum length of a review is set to be 100 words, and reviews shorter than 100 words are padded. The maximum number of reviews is set to be 15. The resulting documents are the input of the embedding layer described in section 3. We use pre-trained word vectors trained on the Google News3 corpus as initiation of our word embeddings.

\subsection{Evaluation Metrics}
To compare our model with other baselines models, we use mean absolute error (MAE). MAE is sometimes preferred over RMSE, due to the fact that RMSE is a function of a set of errors, rather than one average-error magnitude (MAE). Given a predicted rating ${\hat{y}}_{u,i}$ and a ground-truth rating $y_{u,i}$ from the user $u$ for the item $i$, the MAE is calculated as:
\begin{equation}
E_{MAE}= {\frac{1}{N}}{\sum\limits_{u,i} }|{\hat{y}}_{u,i}-y_{u,i}|
\end{equation}
where $ N $ also indicates the number of ratings between users and items.

\subsection{Baselines}
We compare our model with multiple existing models, including matrix factorization methods, topic modeling methods and deep collaborative filtering models combined with textual information. We also compare components of our model with or without the quantum language model layer.

\textbf{Probabilistic Matrix Factorization (PMF)} \cite{mnih2008probabilistic}. PMF is a probabilistic version of traditional matrix factorization methods, which scales linearly with the number of observations and performs well on large, sparse and imbalanced data.

\textbf{Hidden Factors and Hidden Topics (HFT)}\cite{mcauley2013hidden}. HFT combines latent dimensions of ratings with latent dimensions in review topics, which improves both rating prediction and interpretability of recommendation. 

\textbf{Deep Convolutional Neural Network (DeepCoCNN)}\cite{zheng2017joint}. DeepCoCNN uses two one-dimensional convolutional neural networks to learn the latent representation of user and items from all reviews of a user and an item. Factorization Machine was utilized at the top layer of the model to learn interactions between latent representations of user and item.

\textbf{Dual-Attention Convolutional Neural Network (Dual-Att)} \cite{seo2017representation}. Dual-Att adds a local attention layer and a global-attention layer on top of the convolutional neural network to learn a better representation of users and items from reviews. 

\textbf{Transformational Neural Networks (TransNets)}\cite{catherine2017transnets}. TransNets extends the DeepCoNN model by introducing an additional component that learns latent representation of a target user’s review of a target item, and regularizes this latent representation with the latent layer representing the target user and target item pair.

\textbf{Neural Attentional Rating Regression with Review-level
Explanations (NARRE)} \cite{chen2018neural}. NARRE not only predicts precise ratings, but also learns the usefulness of each review simultaneously by introducing a novel attention mechanism on top of CNN representation layers. 
\textbf{Convolutional Quantum Language Model with Mutual Attention for Rating Prediction (ConQAR)}. ConQAR is our model that includes the convolutional layer, quantum density matrix layer and mutual-attention layer. 

\subsection{Experiments Details}
We randomly split the dataset into training (80\%), validation (10\%), and test (10\%) sets. We use the validation set for hyper-parameter tuning and use the test set for final performance comparison. All models' hyper-parameters are tuned on the validation set. For our model, the size of convolutional kernel is searched within [(1),(2),(3),(1,2,3)], the number of kernels to be within [50,100,150], the number of fully connected layers to be within [1,2,3,4], and the $\beta$ for the loss of constraint on user and item density matrix is searched within [0.1,0.3,0.5,0.7,0.9]. The learning rate is searched within [0.1, 0.01, 0.001, 0.0001]. 

\subsection {Experimental Results}
The results of rating prediction of our model and baselines are listed in Table~\ref{tab2} for MAE. From these results, several observations can be made.

First, models that incorporate textual information, such as HFT, TransNets and our model conQAR, outperform collaborative filtering approach PMF that does not incorporate textual information. This is not surprising, since textual information introduces extra information about a user and an item. However, we also observe varied performance on different datasets for Dual-Att and NAREE compared to PMF, where in some cases the two deep models perform better, while in other cases poorer. It is our observation that incorporating textual information into modeling does not always guarantee improvement of rating prediction, and sometimes may even hurt performance by a sub-optimal representation of textual information and introducing more noise. 

Second, the deep models, compared to HFT that utilizes topic modeling to incorporate textual information, have varied performances. DeepCoCNN consistently performs relatively poorly in all datasets. To be fair, the results in the original paper was better \cite{zheng2017joint}. However, when corresponding review information between a user and an item in the test set is excluded, the model performance deteriorates, as reported in \cite{catherine2017transnets}. Dual-Att performs better on larger datasets (Grocery and Gourment Food, Digital Music and Yelp), probably due to that it utilizes textual information alone, but  not independent user and item embeddings for final dot product, therefore the performance of the model is poorer on smaller datasets where textual information is insufficient to learn all parameters in the model.

% \begin{table}
% \centering
% \caption{Performance comparison on eight datasets for all methods (RMSE).}\label{tab2}
% \begin{tabular}{|p{1.3cm}|p{1.3cm}|p{1.3cm}|p{1.3cm}|p{1.3cm}|p{1.3cm}|p{1.3cm}|p{1.3cm}|p{1.3cm}|}
% \hline
% &Musical Instruments&Auto-motive&Amazon Instant Video&Office Products&Patio Lawn and Garden&Grocery and Gourmet Food&Digital Music & Yelp\\
% \hline
% PMF &0.9208 &0.9546 &1.1248 &0.9424 &1.1273 &1.0973 &1.0955 &1.2193\\
% HFT &1.0067 &0.9681 &0.9636 &0.8700 &1.0729 &\textbf{1.0022} &0.9197 &1.1389\\
% DeepCo-CNN &0.9755 &0.9993 &1.2232 &0.9960 &1.1705 &1.1740 &1.1499 &1.4564\\
% Dual-Att &1.2467 &1.0423 &1.0817 &1.0415 &1.2225 &1.0924 &1.0325 &1.2071\\
% TransNet &0.9135 &0.9441 &0.9920 &0.9207 &1.0822 &1.0383 &0.9845 &1.1701\\
% NARRE &0.9491 &0.9910 &1.1947 &0.9830 &1.1759 &1.1249 &1.1349 &1.2886\\\hline
% conQAR &\textbf{0.9028} &\textbf{0.9125} &\textbf{0.9598} &\textbf{0.8695} &\textbf{1.0209} &1.0063 &\textbf{0.9153} &\textbf{1.1298}\\
% \hline
% \end{tabular}
% \end{table}

Third, as shown in Table~\ref{tab2}, our method ConQAR consistently outperforms all the baseline methods. As argued previously, despite the fact that textual information is
useful in recommendation, there is no guarantee that introducing textual information will improve model performance. The performance can vary depending on
how textual information is incorporated into each model effectively. In our model, we propose a
quantum-like density matrix layer on top of convolutional layers to learn covariances between latent features, and a new mutual-attention layer for user and item density matrix to communicate information from counterparts.

\begin{table*}
  \caption{Performance comparison on eight datasets for all methods (MAE).}
  \label{tab2}
  \begin{tabular}{p{1.3cm}p{1.3cm}p{1.3cm}p{1.3cm}p{1.3cm}p{1.3cm}p{1.3cm}p{1.3cm}p{1.3cm}}
    \toprule
    &Musical Instruments&Automo-tive&Amazon Instant Video&Office Products&Patio Lawn and Garden&Grocery and Gourmet Food&Digital Music & Yelp\\
    \midrule
    PMF&0.6800&0.7118&0.8890&0.7459&0.8787&0.8741&0.8558&0.9692\\
\hline
HFT&0.7121&0.6792&0.6950&0.6650&0.8142&0.7363&0.6903&0.8869\\
\hline
DeepCo-CNN&0.7256&0.7225&0.9527&0.7653&0.9205&0.9020&0.8856&1.1641\\
NARRE&0.7049&0.7092&0.9156&0.7964&0.9672&0.8776&0.9030&1.0738\\
Dual-Att&1.0864&0.7647&0.8700&0.8695&1.0097&0.8016&0.7743&0.9640\\
TransNet&0.6370&0.6073&0.6888&0.6766&0.8147&0.7047&0.6975&0.8994\\
\hline
ConQAR&\textbf{0.6031}&\textbf{0.6046}&\textbf{0.6620}&\textbf{0.6286}&\textbf{0.7289}&\textbf{0.6997}&\textbf{0.6449}&\textbf{0.8717}\\
    \bottomrule
  \end{tabular}
\end{table*}

\section {Ablation Test}
To analyze how each component affects the performance of the model, we conducted an ablation test. We compare two variants of the proposed model with the full model. The first variant is composed of a convolutional layer and a mutual-attention layer (Conv+MutualAtt), in which case the mutual-attention matrix is the multiplication of two feature maps between user and item. The second variant includes a convolutional layer and a quantum density matrix layer, and uses the same final representation as in \cite{zhang2018end}, which uses the trace and diagonal elements of the multiplication of density matrices as final representations of a user and an item (Conv+Quant). The full model includes a convolutional layer, a quantum density matrix layer and a mutual-attention layer (Conv+Quant+MutualAtt). We compare the performances of the variants on MAE.

The results of the test are listed in table ~\ref{tab3}. From the table, we can observe that both components, namely quantum-like density matrix layer and the mutual-attention layer have improved model performance over the baseline model, namely the DeepCoCNN model \cite{zheng2017joint}, which only has the convolutional layers. 

Specifically, by adding the quantum-like density matrix layer alone, the performance of our model has improved considerably. This is also observed in previous work in quantum-like language modeling with application in information retrieval\cite{sordoni2013modeling} and question answering \cite{zhang2018end}, that by introducing the quantum-like language model, extra information of feature-interaction is incorporated into the model that was neglected in previous model. 

By adding the mutual-attention layer independently, the performance is significantly improved. We speculate that this is probably due to the fact that mutually-shared information between a user and an item, embedded in high-level n-gram representations (be it CNN features or quantum-like density matrices) are given more weight through the mutual-attention layer,  therefore making the final rating prediction more relevant to the specific context of this user-item pair. Finally, by incorporating both the quantum-like density matrix layer and the mutual-attention layer, the performance is further improved over each variant and the baseline model.

\begin{table*}
  \caption{Performance comparison on six datasets for ablation test (MAE).}
  \label{tab3}
  \begin{tabular}{p{3.8cm}p{1.3cm}p{1.3cm}p{1.3cm}p{1.3cm}p{1.3cm}p{1.3cm}}
    \toprule
    &Auto-motive&Amazon Instant Video&Office Products&Patio Lawn and Garden&Grocery and Gourmet Food&Digital Music\\
    \midrule
    DeepCo-CNN&0.7256&0.9527&0.7653&0.9205&0.9020&0.8856\\
\hline
Conv+Quant &0.6507&0.6854&0.6440&0.7728&0.7550&0.6767\\
Conv+Mutual&0.6088&0.6781&0.6355&0.7454&0.7245&0.6619\\
Conv+Quant+MutualAtt&\textbf{0.6046}&\textbf{0.6620}&\textbf{0.6286}&\textbf{0.7289}&\textbf{0.6997}&\textbf{0.6449}\\
    \bottomrule
  \end{tabular}
\end{table*}

\section {Visualization of Density Matrices}

In previous section, we argue that one-dimensional convolutions may be insufficient for the representation of user and item reviews, since the covariance information between latent dimensions of feature vectors in convolutional feature maps are not well captured in previous models. 
Some models tried to capture such interactions between latent dimensions of feature vectors with fully-connected layers or factorization machine \cite{zhang2018end,catherine2017transnets}. However, such interactions are learned at very late stage of the model, where the feature maps are already max-pooled or mean-pooled into one-dimensional vectors. Instead, we propose to apply a quantum-like density matrix layer on top of the convolutional layer, where such interactions of the latent dimension of feature vectors are well captured by dyads of unit states and weighted into a density matrix.

\begin{figure}
\includegraphics[width = 0.4\textwidth]{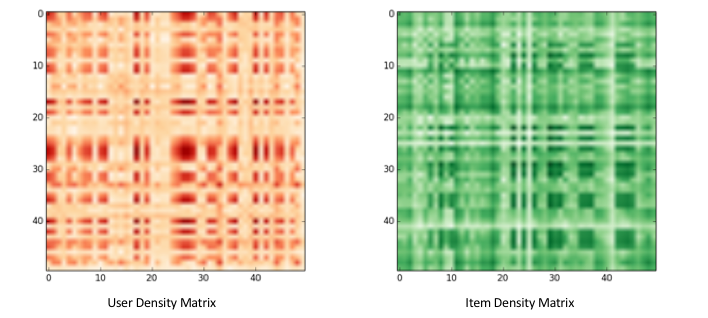}
\caption{Visualization of user and item density matrix.Left: density matrix for one user. Right: density matrix for one item. Number of latent features=50. Darker shades suggest higher values in density matrices (higer co-variance).} \label{fig2}
\end{figure}

To investigate whether such density matrices indeed capture interactions between latent dimensions of feature vectors, we visualize the density matrices of a user and an item that are randomly selected. The visualization is illustrated in Fig.~\ref{fig2}. The user (left) and item (right) density matrices are depicted with two $50 \times 50$ heatmaps for 50 filters, where darker shades indicate greater values in density matrices.

From the visualizations, we can observe that there are high covariances among latent features of a particular user and item, indicated by the darker areas in both heat maps. This co-variance further relates similar features (n-grams) together, and promotes communication between these similar features at later mutual-attention layer. One-dimensional convolutional layers, on the other hand, may fail to capture such relations by applying pooling at early stage of the models.

\begin{figure}
\centering
\begin{subfigure}
  \centering
  \includegraphics[width=\linewidth]{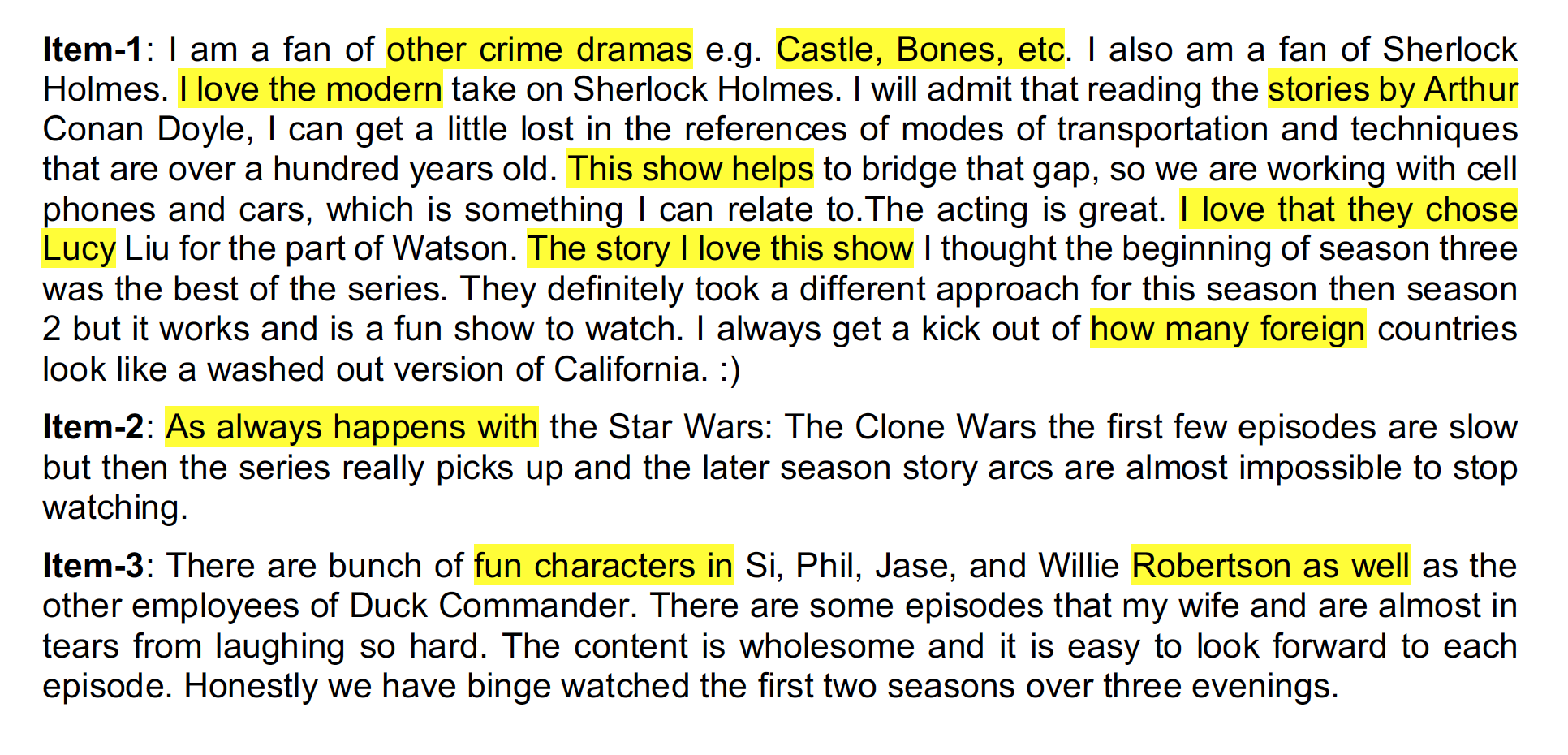}
\end{subfigure}%
\begin{subfigure}
  \centering
  \includegraphics[width=\linewidth]{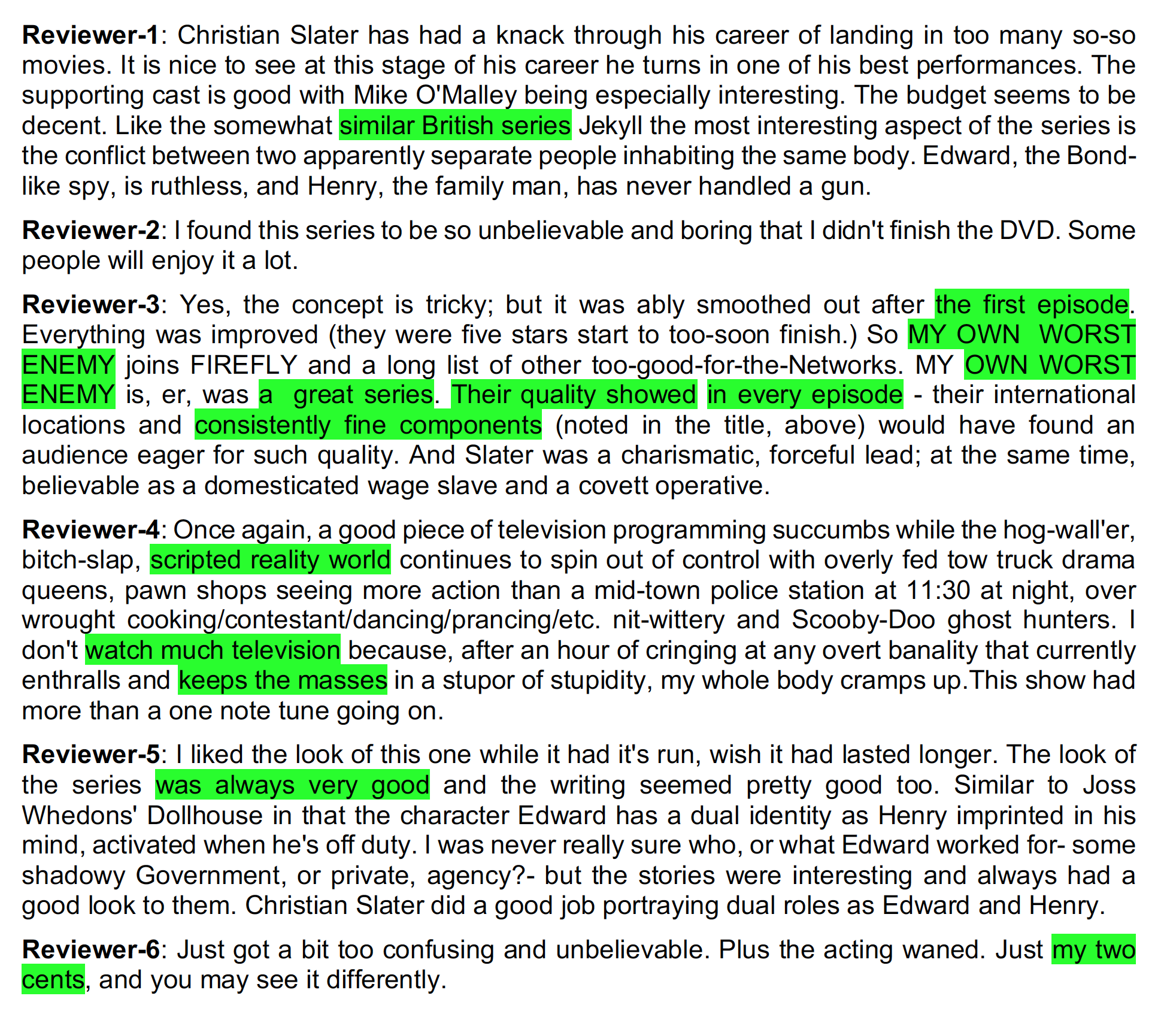}
\end{subfigure}
\caption{Visualization of Attentions of N-grams at different positions of review document of user and item. Top: one user review document. Bottom: one item review document.}
\label{fig3}
\end{figure}

\section {Visualization of Attention Over N-grams in User and Item Reviews}
As described earlier, $\vec{p}_u$ ($\vec{p}_v$) is the probabilistic distribution over states of n-grams of a review document for a user $u$ or an item $v$. This distribution determines the final density matrix as a mixture of elementary dyads. 

Since the true probabilistic distribution space over all possible n-gram combinations are infinitely large, we approximate the distribution with probabilistic distribution over all positions of a document, and consider such distributions as user and item-specific. 
To investigate whether $\vec{p}_u$ ($\vec{p}_v$) learns anything meaningful, we visualize the corresponding review document of a user and an item, and highlight each position according to values in $\vec{p}_u$ ($\vec{p}_i$). Since $p_u$ ($p_i$) is user/item-specific but not pair-specific, therefore it can be considered as overall profiling of a user and item, by looking at the more important positions in the review document of the user or item according to values in $\vec{p}_u$ ($\vec{p}_i$).

As visualized in Fig.~\ref{fig3}, we visualize a review document of a user A and an item B from the category of Amazon Instant Videos, where the top 20 positions in the review document are highlighted with yellow and green colors. The positions are ranked by the corresponding values in $\vec{p}_u$ and $\vec{p}_i$. Note that each user and item review document are comprised of multiple reviews of items and users. 

User A has 4 reviews on 4 different videos (Fig.~\ref{fig2}-top). It can be observed that the model puts higher probabilities on genre preferences, such as \textquotedblleft other crime dramas\textquotedblright, \textquotedblleft stories by Arthur\textquotedblright, and also on sentiments such as \textquotedblleft I love the modern\textquotedblright, \textquotedblleft I love that they chose Lucy\textquotedblright, \textquotedblleft The story I love this show\textquotedblright, \textquotedblleft fun characters\textquotedblright. This helps to retain key information about this user's preferences in the final density matrix. It is also interesting to see that the learned probabilities seem to be able to recover similar n-grams with similar probabilities (higher probabilities in this case), for example for the \textquotedblleft I love-\textquotedblright phrases. This may be considered as evidence that our approximation of the true distribution over all possible n-gram combination is valid and can recover similar n-grams with similar probabilities.

Item B is reviewed by 6 different users (Fig.~\ref{fig3}-bottom). From the visualization, we can identify some commentary n-grams from different users about this video, such as \textquotedblleft similar British series\textquotedblright, \textquotedblleft a great series\textquotedblright, \textquotedblleft their quality showed in every episodes\textquotedblright, \textquotedblleft consistently fine components\textquotedblright, \textquotedblleft scripted reality world\textquotedblright, \textquotedblleft was always very good\textquotedblright. We also observe the same pattern that similar or identical n-grams are consistently given similar probabilities, such as the \textquotedblleft MY OWN WORST ENEMY" phrases.

% \begin{figure}
% \centering
% \begin{tabular}{@{}c@{}}
%   \includegraphics[width=1\linewidth]{user_final.png}
% \end{tabular}
% \begin{tabular}{@{}c@{}}
%   \includegraphics[width=1\linewidth]{item_final.png}
% \end{tabular}
% \caption{Visualization of Attentions of N-grams at different positions of review document of user and item. Top: one user review document. Bottom: one item review document. } \label{fig2}
% \end{figure}

\section{conclusion}
In this paper, we propose a convolutional quantum-like language model with mutual-attention for product rating prediction. The motivations behind the proposed model are two-fold. First, most existing CNN-based neural models for product review representation perform max/mean-pooling right after one-dimensional convolutions, where the interaction information between latent dimensions of a feature vector is lost along the way. To bridge this gap, we propose to add a quantum-like density matrix layer on top of the convolutional layer, so that covariances of latent dimensions are well captured in dyads and density matrices. Second, most existing models combine the representations of a user and an item with vector dot-product, vector concatenation plus fully-connected layers or factorization machine, which are not directly applicable for 2-dimensional density matrices in our model. Also, the direct combination takes all features as equal, and does not emphasize features that stand out in both user and item. Instead, in our model, we incorporate both the matrix \textquotedblleft dot-product \textquotedblright, namely the trace of matrix multiplication, and also weighted vectors by attention from the mutual-attention layers. 

Experiments on two large datasets show that our model outperforms all strong baselines. Moreover, the results of the ablation test suggest that both quantum density matrix layers and mutual-attention layers are beneficial for improving model performance. We also present two cases of visualizations, to show that the density matrix layers capture covariances of latent dimensions, and the probabilistic distributions learned in our model are informative and meaningful and capture important information for each user and item.

The present paper also has limitations. We evaluate our model only on metric of MAE, due to the fact that nDCG values of most methods are very similar since the 5-star and 4-star samples are dominant in the Amazon and Yelp datasets, therefore most baselines and our model achieve similar nDCG of over 95\%. In the future, we plan to experiment with our model on more diverse datasets. We would also like to explore to apply the quantum-language model to other base neural networks such as RNNs\cite{cho2014learning}, Transformer \cite{vaswani2017attention}, Bert \cite{devlin2018bert}, to see whether the quantum-language model help to further improve state-of-the-art language models.

\bibliographystyle{ACM-Reference-Format}
\bibliography{sample-base}

\end{document}